\documentclass[journal,10pt]{IEEEtran}
\usepackage{amsmath,amssymb,amsfonts,mathrsfs,bm}
\usepackage{amstext}
\usepackage{upgreek}
\usepackage{multicol}
\usepackage{indentfirst}
\usepackage{graphicx}
\usepackage{paralist}
\usepackage{hyperref}
\usepackage[numbers,sort&compress]{natbib}
\usepackage{booktabs}
\usepackage{multirow}
\newtheorem{proposition}{Proposition}

\IEEEoverridecommandlockouts

\begin{document}
\title{High Throughput Opportunistic Cooperative Device-to-Device Communications With Caching}
\author{Binqiang Chen, Chenyang Yang and Gang Wang
%
\thanks{
	Binqiang Chen and Chenyang Yang are with the School of Electronics and Information Engineering, Beihang University, Beijing, China, Emails: \{chenbq,cyyang\}@buaa.edu.cn. Gang Wang is with the NEC labs, China, Email: wang\_gang@nec.cn.
	
}
}

\maketitle

\begin{abstract}
To achieve the potential in providing high throughput for cellular networks by device-to-device (D2D) communications, the interference among D2D links should be carefully managed. In this paper, we propose an opportunistic cooperation strategy for  D2D transmission by exploiting the caching capability at the users to control the interference among D2D links. We consider overlay inband D2D, divide the D2D users into clusters, and assign different frequency bands to cooperative and non-cooperative D2D links. To provide high opportunity for cooperative transmission, we introduce a caching policy. To maximize the network throughput, we jointly optimize the cluster size and bandwidth allocation, where the closed-form expression of the bandwidth allocation factor is obtained. Simulation results demonstrate that the proposed strategy can provide $400\% \sim 500\%$ throughput gain over traditional D2D communications when the content popularity distribution is skewed, and can provide $60\% \sim 80\%$ gain even when the content popularity distribution is uniform.
\end{abstract}


\section{Introduction}

Device-to-device (D2D) communications enables direct communications between two user
devices without traversing the base station (BS) or core network, and is a promising way to achieve the high throughput goal of 5th generation (5G) cellular networks \cite{doppler2009device,Andrews.D2D,zhang2013exploring,Andreev.JSAC}. The typical use-cases of D2D communications include
cellular offloading, content distribution, and relaying, \emph{etc.} \cite{Survey.D2D}, where content delivery service has attracted considerable attention recently, since it accounts for the majority of the explosive increasing traffic load.

Motivated by the observation that a large amount of content delivery requests are asynchronous but redundant, i.e., the same content is requested repeatedly at different times, caching has long been studied as a technique to improve performance of wired networks. Due to the rapid reduction in cost of storage device, caching at the wireless edge is also recognized as a promising way for delivering popular contents nowadays, which can improve the network throughput, energy efficiency and the quality of user experience (QoE) \cite{wang2014cache,Procach14,Ali13,Dong,Higgins12,LHui14,Chen15}. However, different from wired networks, the performance of wireless networks is fundamentally limited by the interference, which inevitably limits the throughput gain from local caching.

To take the advantage of the storage device at smart phones, cache-enabled D2D communications has been proposed recently, which can offload the content delivery traffic and hence boost the network throughput significantly \cite{Golrezaei.TWC,JMY.JSAC}. Since only the users in proximity communicate to each other, the interference in D2D networks is strong, which needs to be carefully controlled.
In an early work of studying cache-enabled D2D communications, the D2D users are divided into clusters. Then, the intra-cluster interference among D2D links is managed by using time division multiple access (TDMA), while the inter-cluster interference between D2D links is simply treated as noise \cite{Golrezaei.TWC}. In \cite{JMY.JSAC}, only the D2D link from one of the four adjacent clusters is allowed to be active at the same time-frequency resource block, in order to avoid strong inter-cluster interference among adjacent clusters.
In \cite{IA.D2D}, interference alignment was employed to  mitigate the interference  among D2D links, but only three D2D links were coordinated within each cluster, and the interference among clusters was again treated as noise.
In \cite{CD2D.15,GC13.Relay,C2D.Relay}, cooperative relay techniques were proposed to mitigate the interference between cellular
and D2D links, which however can not manage the interference among the D2D links.

It is well known that if several transmitters have the required data for some users, they can jointly transmit to the users without generating interference. In fact, if contents have been locally cached, cooperative transmission without data exchange among transmitters becomes possible, which can transform interference into spatial multiplexing gain. Based on such an interesting observation, a BS cooperative transmission strategy was proposed in \cite{Lau.Tran13} by exploiting the caches at BSs, where precoding and cache control were optimized to guarantee the QoE of users. Inspired by this work, a natural question is: can we apply cooperative transmission in D2D communications with caching?

Fortunately, cooperative transmission is possible in practice due to the following reasons. (i) In D2D communications, the D2D transmitter (DT) has been proposed to assist other users in additional to transmitting data to its destined D2D receiver (DR), e.g., with cooperative relay \cite{GC13.Relay}. The users have the incentive to do this if their own QoE can be improved or their costs can be compensated by some other rewards \cite{incentive}. (ii) To facilitate cooperative transmission, the global channel state information (CSI) is required to compute the precoding matrix. The CSI among D2D links can be obtained at DTs and the BS through channel probing and feedback \cite{CSI}. Then, the precoding vectors can be computed at the BS and sent to the cooperative DTs via multicast. (iii) The synchronization among cooperative DTs is more easier to be implemented than that in Ad-hoc networks, because it can be realized with the assist of the BS \cite{Andrews.D2D}. Besides, the synchronization can also be realized at users by using the methods proposed in \cite{SYN.13}.

In this paper, we propose an {opportunistic cooperation strategy} for cache-enabled D2D communications to manage the interference among D2D links. Different from the BS cooperative transmission strategy \cite{Lau.Tran13}, the cooperation strategy for D2D communications needs to be optimized in a different way. Considering that D2D communications is applicable to users in proximity, we divide the D2D users into virtual clusters. To maximize the opportunity of cooperative transmission via D2D links, we take both redundant caching and diversity caching into account in the users among the clusters, which differs from \cite{Lau.Tran13} where all BSs cache the
same files. When some users have cached the files requested
by other users called DRs, these users act as DTs to
jointly transmit the requested files to the DRs.
Because only some D2D links can employ cooperative transmission, we assign different frequency bands to cooperative and non-cooperative links to avoid mutual interference. To maximize the average network throughput without compromising the experience of non-cooperative users, we jointly optimize the cluster size and bandwidth allocation under the minimal average user data rate constraint.

The contributions of this paper are summarized as follows:
\begin{itemize}
	\item We propose an opportunistic cooperation strategy to manage the interference among D2D links, which improve the network throughput remakably.
	\item We jointly optimize the cluster size and bandwidth allocation and obtain the closed-form expression of optimal bandwidth allocation factor.
\end{itemize}

The rest of the paper is organized as follows. Section \ref{sec:system model} presents the system model. Section \ref{sec:optimiazation} introduces the cooperation strategy, derives the average network throughput and average user data rate, and jointly optimizes bandwidth allocation and cluster size. Section \ref{sec:simulation} provides numerical and simulation results. Section \ref{sec:conclusion} concludes the paper.

\section{System Model}
\label{sec:system model}

\begin{figure}[!hb]
	\centering
	\includegraphics[width=0.45\textwidth]{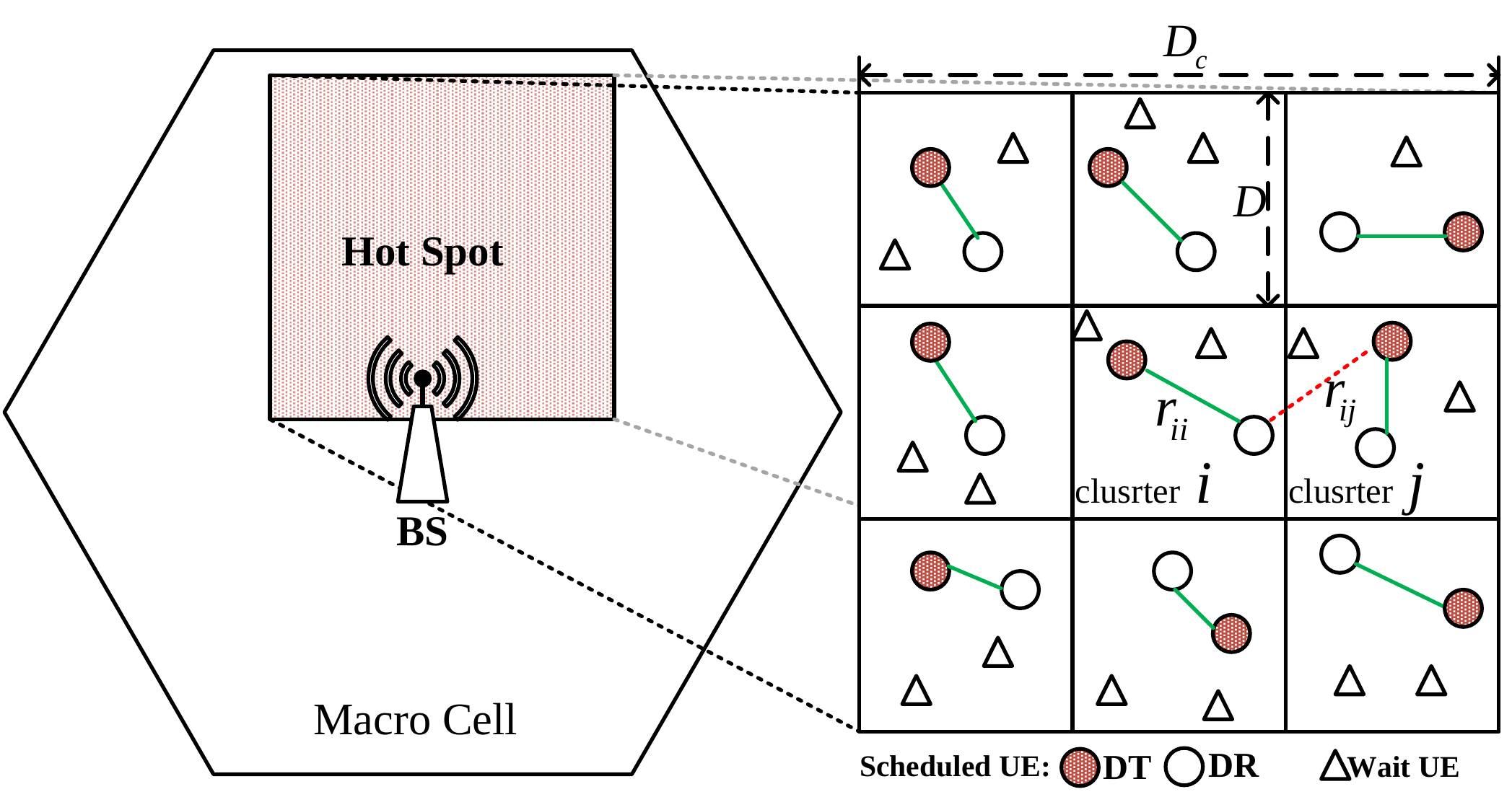}\\
	\caption{Cluster division model, ``UE" means user equipment.}\label{fig.ici-d2d}
\end{figure}

Consider a cellular network, where $M$ single-antenna users are uniformly located in a square hotspot within a macro cell,
where the area is with side length of $D_c$ as shown in Fig. \ref{fig.ici-d2d}. Each user is willing to store $N$ files in its local cache and can act as a helper to share files. When a helper conveys a file in local cache via D2D link to a DR requesting the file, the helper becomes a DT. The BS is aware of the cached files at each user and coordinates the D2D communications.

\subsection{Content Popularity}

We consider a static content catalog including $N^f$ files that the users may request, where the files are indexed in a descending order of popularity, e.g., the 1st file
is the most popular file. The probability that the $i$th file is requested by a user is assumed to follow a Zipf distribution,
\begin{equation}
	\label{equ.P_N_f}
	P_{N^f}(i)=i^{-\beta}/\sum_{j=1}^{N^f}j^{-\beta},
\end{equation}
where $\sum_{i=1}^{N_f}P_{N^f}(i)=1$, and the parameter $\beta$ reflects skewness of the popularity distribution, with large $\beta$ meaning that a few files are requested by the majority of users \cite{Zipf99}.

\subsection{Communication Protocol}
D2D links can be established among users in proximity. A widely used communication protocol for D2D communications is that two user equipments (UEs) can communicate if their distance is smaller than a given distance \cite{Golrezaei.TWC,Kumar.C}.
To restrict the D2D link distance and make the analysis tractable, the square hotspot area is divided into $B$ smaller square areas called clusters, where the side length of each cluster is $D=D_c/\sqrt{B}$. Only the users within the same cluster can establish D2D link. For mathematical simplicity, we assume that the number of users per cluster is $K=M/B$ and each user is assumed to transmit with the same power $P$ as in \cite{JMY.JSAC}.

We consider overlay inband D2D \cite{Survey.D2D}, and assume that a fixed bandwidth of $W$ is assigned to the D2D links.

\section{Opportunistic Cooperation Strategy}
\label{sec:optimiazation}
In this section, we first introduce a caching policy to provide high opportunity for cache-enabled cooperative D2D transmission. Then, we propose an opportunistic cooperative transmission policy. Finally, we optimize two key parameters in the strategy to maximize the network throughput.

\subsection{Caching Policy}
To maximize the probability that a user can fetch files through D2D links, the users within a cluster should cache different files. To maximize the probability of cooperative transmission among DTs in different clusters, the files cached at the users of each cluster should be the same.
This suggest that the caching policy needs to balance the
diversity of content with the redundancy of the replicas of popular
contents. To this end, we consider the following caching policy.

According to the user cache size $N$, all files are divided into $K_0=N^f/N$ groups. The $k$th \emph{file group} $\mathcal{G}_k$ consists of the $(k-1)N+1$th to the $kN$th files where $1 \leq k\leq K_0$, e.g., the 1st file group $\mathcal{G}_1$ contains the most popular $N$ files. Then, the probability that a user requests a file within the $k$th \emph{file group} $\mathcal{G}_k$ can be obtained as
\begin{equation}
	\label{equ.P_k}
	P_k= \sum_{i=(k-1)N+1}^{kN}P_{N^f}(i) = \frac{\sum_{i=(k-1)N+1}^{kN}i^{-\beta}}{\sum_{j=1}^{N^f}j^{-\beta}} .
\end{equation}

In every cluster, the $k$th user caches the $k$th file group $\mathcal{G}_k$. Then, every user in each cluster caches different files, i.e., diversity caching is achieved within each cluster, and the most popular $KN$ files are cached in every cluster, i.e., redundant caching is achieved among clusters. Because each cluster contains $K$ users, the file groups with indices exceeding $K$ (i.e.,  $\mathcal{G}_k$, $k>K$ ) are not cached at users.

When $K=K_0$, all the $N^f$ files can be cached at the users in each cluster and all user requests can be served via D2D links, therefore it is not necessary to assign more than $K_0$ users to each cluster. For this reason, we assume $K \leq	K_0$.

In practice, the files can be proactively downloaded by the operator from the  BS to the cache at each user via multicast during off-peak times according to the  user demand statistics.


\subsection{Opportunistic Cooperative Transmission Policy}
According to whether a user can find the requested file in its local cluster, we can classify the users into two types.

{\bf D2D users:} If the file requested by a user is cached at any UE in the cluster it belongs to (called \emph{local cluster of the user}), then the user can directly obtain the file with D2D communication. Such a user is referred to as a D2D user. Besides, if the file requested by a user is in its local cache, it can retrieve the file immediately with zero delay, but we ignore this case for analysis simplicity as in \cite{JMY.JSAC}.

{\bf Cellular users:} If the file requested by a user is not cached in the UEs within its local cluster, the user fetches the file from the BS and becomes a regular cellular user. The number of cellular users is denoted as $N^b$.

\begin{figure}[!h]
	\centering
	\includegraphics[width=0.4\textwidth]{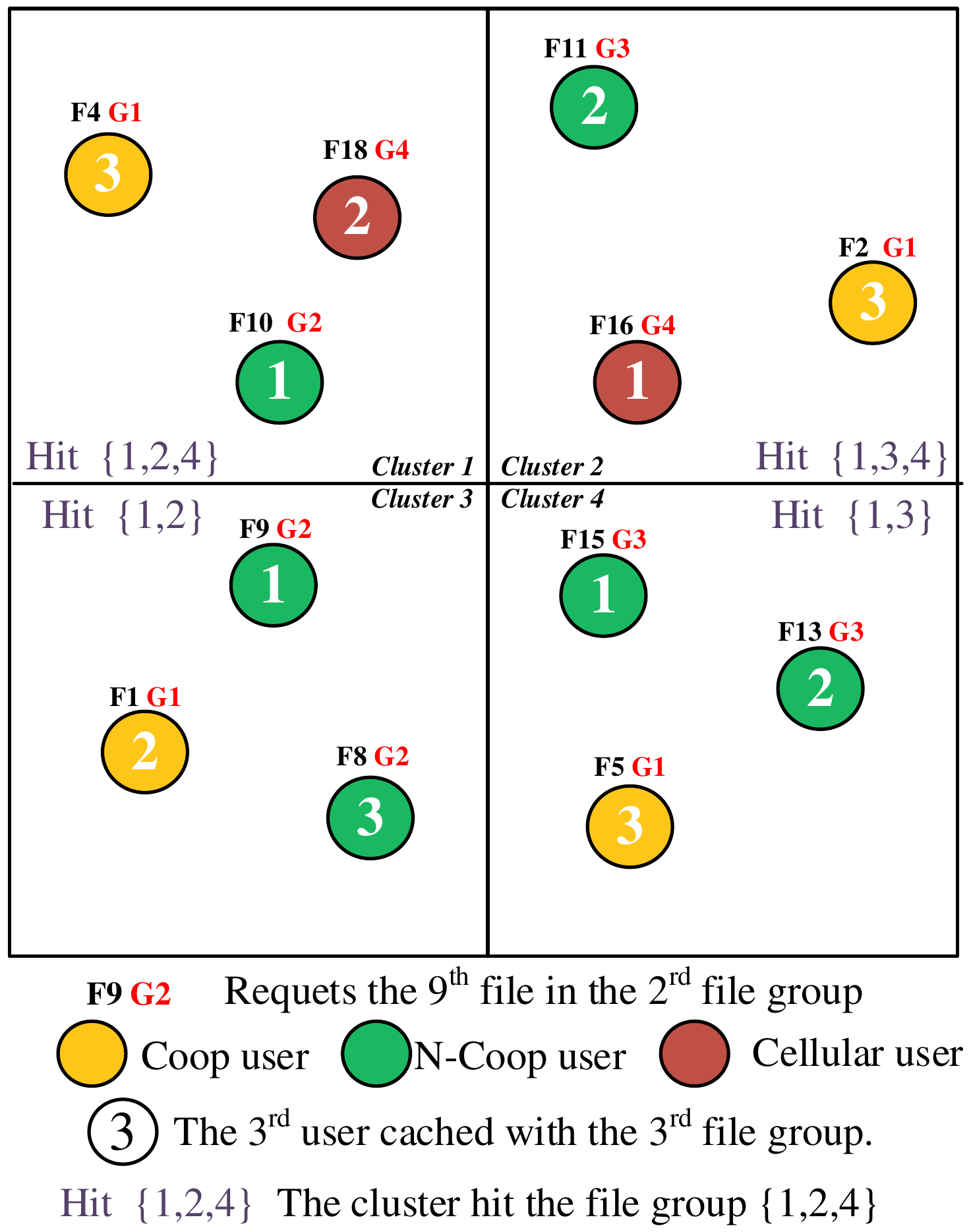}\\
	\caption{Illustration of the opportunistic cooperaton strategy. Catalog size $N_f=20$, $B=4$ clusters in the hotspot, $K=3$ users in each cluster and each user caches $N=5$ files. All $20$ files are divided into $4$ groups according to content popularity, e.g., $\mathcal{G}_1 = \{1,2,3,4,5\}$.}\label{fig.Coop}
\end{figure}

For easy understanding, we introduce the strategy with the help of an example.

\subsubsection{Cooperative D2D Users}
If there exists at least one user in a cluster requesting the files in $\mathcal{G}_k$, then we say that the cluster \emph{hits the $k$th file group}. In Fig. \ref{fig.Coop}, the users in the first cluster respectively request the files in $\mathcal{G}_1$, $\mathcal{G}_2$ and $\mathcal{G}_4$, and hence the first cluster hits the  $\{1,2,4\}$th file groups.

If every cluster hits the same file group $\mathcal{G}_k$, the $k$th user in each cluster who caches the file group $\mathcal{G}_k$ can act as a DT, and all DTs in these clusters  cooperatively transmit files to the DRs requesting the files in $\mathcal{G}_k$.\footnote{Though further improvement is possible by allowing cooperation among less than $B$ clusters (called partial cooperation), we only consider the full cooperation among all clusters for mathematical tractability. The impact of partial cooperation is shown via simulation in Section \ref{sec:simulation}.} Those DRs are referred to as {\bf cooperative D2D users} (\emph{Coop users} for short), whose number is denoted as $N^c$.

\begin{figure}[!htp]
	\centering
	\includegraphics[width=0.4\textwidth]{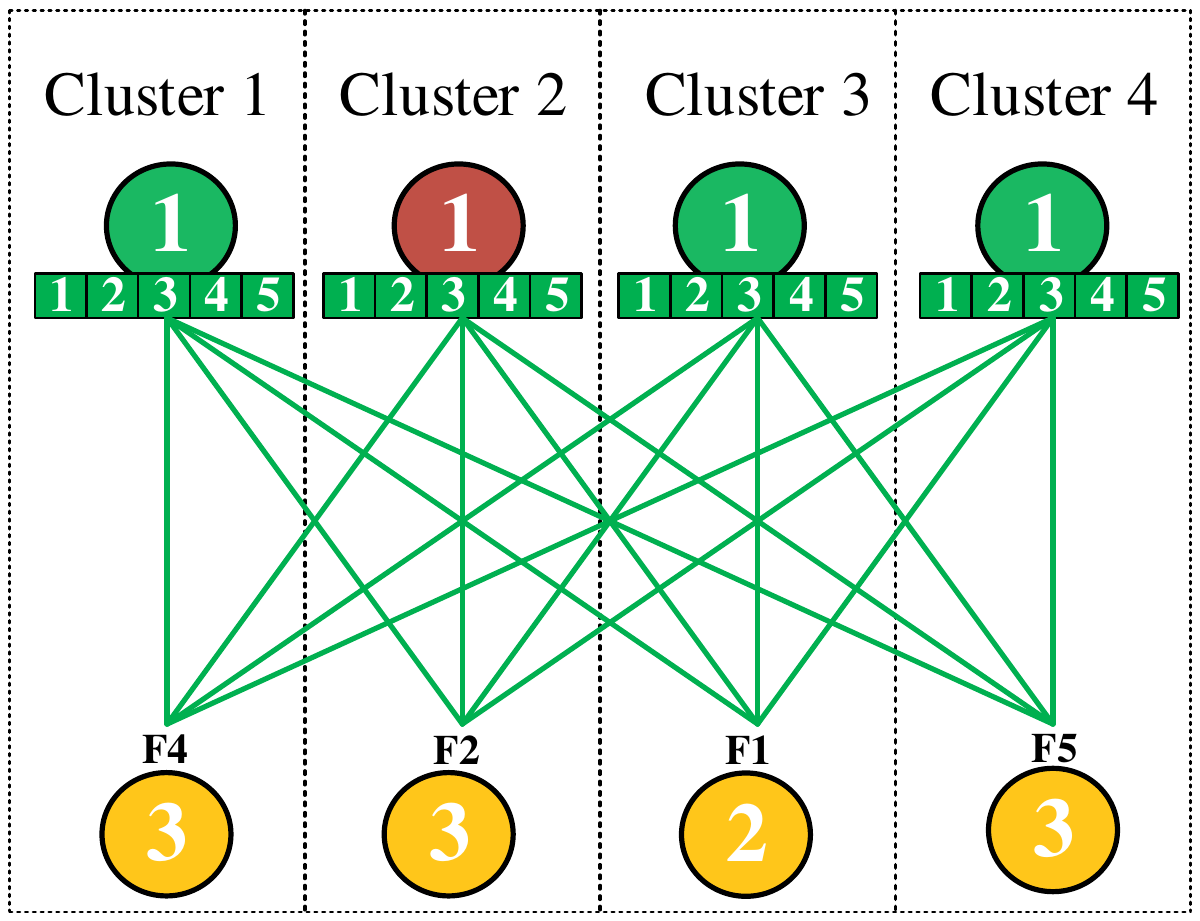}\\
	\caption{Illustration for cooperative transmission from multiple DTs to DRs.}\label{fig.Coop_transmission}
\end{figure}

In Fig. \ref{fig.Coop}, every cluster hits the $1$st file group. Hence, the $1$st users  in all the four clusters who cache the files in $\mathcal{G}_1$ can act as DTs to cooperatively transmit files with indices $\{4,2,1,5\}$ respectively to the $3$rd user in cluster $1$, the $3$rd user in cluster $2$, the $2$nd user in cluster $3$, and the $3$rd user in cluster $4$, as shown in Fig. \ref{fig.Coop_transmission}.

The remaining users except the cellular and Coop users are {\bf non-cooperative D2D users} (\emph{N-Coop users} for short), whose number is $N^n= M-N^b-N^c$.

\subsubsection{Interference Control}
Due to the random locations of the DTs in proximity, the interference in the network needs to be carefully controlled even with the cooperative transmission.

{\bf Inter-type interference:} To avoid the mutual interference between \emph{Coop users} and \emph{N-Coop users}, we assign $\eta W$ for Coop users and the remaining bandwidth $(1-\eta) W$ for N-Coop users, where $\eta$ is the bandwidth allocation factor and $0 \leq \eta < 1$. A large value of $\eta$ means that more bandwidth is allocated to Coop users.

{\bf Intra-cluster interference:} Considering that the users within each cluster can not cooperate due to caching different files, we randomly select one Coop D2D link and one N-Coop D2D link respectively in each cluster to transmit at the same time to avoid intra-cluster interference as in \cite{Golrezaei.TWC,JMY.JSAC}.

{\bf Inter-cluster interference:} There is no inter-cluster interference among Coop users owing to the joint transmission from multiple DTs, but there exists inter-cluster interference between N-Coop users, which is regarded as noise.

\subsubsection{Operation Modes}

Due to the opportunistic nature of establishing the cooperative D2D links, the network may operate in the following two modes.
\begin{itemize}
	\item In \emph{Mode} $0$, there does not exist any file group hit by every cluster, i.e., all D2D users are N-Coop users. Then, all the DTs transmit independently, and the bandwidth $W$ is assigned to the N-Coop users, i.e.,
	$\eta=0$.
	\item In \emph{Mode} $1$, there exist file groups hit by every cluster, i.e., there exist Coop users. Then, $0<\eta<1$.
\end{itemize}

To become the Coop users, the users are not necessary to request the \emph{same file}, but to request the \emph{files in the same group}. Hence the cooperative probability, i.e., the  probability that the network operates in \emph{Mode 1}, is high, which increases with the number of users in each cluster $K$.
To see this, we derive  the cooperative probability $P^c$ as follows.

A cluster hits the $k$th file group if at least one of the $K$ users in the cluster requests a file in $\mathcal{G}_k$, whose probability is denoted as $P^h_k$. It is the complement of the probability that no user requests any file in the $k$th file group, which is $(1-P_k)^K$. Then, from (\ref{equ.P_k}), $P^h_k$ can be obtained as
\begin{equation}
	\label{equ.P^h_k}
	P^h_k=1-(1-P_k)^K
\end{equation}
which  increases with  $K$, because $0 \leq P_k \leq 1$.

Cooperative probability is the probability that there exists at least one file group hit by all the $B$ clusters. It is the complement of the probability that there is no file group hit by all the $B$ clusters, and hence can be derived
as
\begin{equation}
	\label{equ.P^c}
	P^c=1-\prod_{k=1}^{K}(1-(P^h_k)^B),
\end{equation}
where  $1-(P^h_k)^B$ is the probability that the number of clusters hitting the $k$th file group is less than $B$, which decreases with the growth of $K$ since $B=M/K$. Therefore, for a given value of $M$, $P^c$ is an increasing function of $K$.

\subsubsection{Key Parameters}
Since only one Coop D2D link per cluster is allowed to be active each time,  $B$ users out of all Coop users can be scheduled simultaneously in \emph{Mode} 1. Therefore, the number of {\bf active Coop users} is $N^a=B$ in \emph{Mode} $1$, and is $N^a=0$ in \emph{Mode} 0.\footnote{For the considered strategy, the number of active Coop users is less than the number of Coop users, i.e., $N^a \le N^c$.} With the cooperative probability, the average number of active Coop users can be obtained as
\begin{equation}
	\label{equ.E_N^a}
	\bar{N^a}= BP^c+0(1-P^c)=BP^c.
\end{equation}

$\bar{N^a}$ characterizes how many \emph{interference-free D2D links} can transmit at the same time-frequency resources in average, which can reflect  the multiplexing gain. In general, the number of interference-free D2D links demonstrates the same trend with the network throughput, as to be verified in Section \ref{sec:simulation}. Hence, a large value of $\bar{N^a}$ implies a high network throughput. When the number of users per cluster $K$ is large, the cooperative probability is high, but the number of active Coop users is small since $N^a=B$ and $B=M/K$. This suggests that there is a tradeoff between two counter-running effects: a small value of $K$ leads to more active Coop users if the system operates in Mode 1; a large value of $K$ yields high cooperative probability. In other words, to maximize the network throughput, the cluster size should be optimized, which is reflected by the number of users per cluster $K$ since the number of users in the hotspot $M$ is given.

Due to the multiplexing gain and interference-free transmission, the average data rate of Coop users usually exceeds that of N-Coop users. As a result, the overall network throughput will be reduced if we simply assign identical bandwidth to these two types of D2D users. Owing to the same reason, simply allocating all the bandwidth to Coop users can maximize the network throughput, but no N-Coop users can be served. This indicates that the bandwidth allocation factor $\eta$ should be optimized to maximize the throughput of the network under the constraint on the data rate of each user to avoid unfairness.

\subsection{Optimization of Cluster Size and Bandwidth Allocation}
In this subsection, we jointly optimize the bandwidth allocation factor $\eta$ and cluster size $K$ to maximize the average network throughput under a constraint that the average user data rate is larger than a given value, $\mu$ (Mbps).
Because we assume overlay D2D communications, only D2D users are considered in the network throughput.

In the sequel, we first derive the average network throughput achieved by all D2D users. Then, we derive the average data rate for each of Coop and N-Coop users. Finally, we find the optimal cluster size $K$ and bandwidth allocation factor $\eta$.

\subsubsection{Average Network Throughput}
Recall that only one Coop D2D link (if any) and one N-Coop D2D link are scheduled per cluster in each time. Then,
the average throughput of the network operating in \emph{Mode} 0 can be obtained as follows,
\begin{equation}
	\label{equ.bar_R_0}
	\bar{R_0} = \mathbb{E}\{W\sum_{i=1}^{B}R_i^n\} \overset{(a)}{=} WB\bar{R_i^n},
\end{equation}
where the expectation is taken over small scale channel fading and user location, $(a)$ comes from the fact that all users are randomly located and transmit with equal power, and ${R_i^n}$ and $\bar{R_i^n}$ are the instantaneous and average data rate per unit bandwidth per second  of the N-Coop link in the $i$th cluster,  respectively.

Analogically, the average throughput of the network operating in \emph{Mode} 1 can be obtained as
\begin{equation}
	\label{equ.bar_R_1}
	\begin{split}
		\bar{R_1} &= \mathbb{E}\{ \eta W\sum_{i=1}^{B}R_i^c\ + (1-\eta)W\sum_{i=1}^{B}R_i^n\} \\
		&= WB(\eta \bar{R_i^c}+(1-\eta) \bar{R_i^n} ),
	\end{split}
\end{equation}
where ${R_i^c}$ and $\bar{R_i^c}$ are the instantaneous and average data rate per unit bandwidth per second of the Coop link in the $i$th cluster,  respectively.

Further considering the cooperative probability $P^c$ in (\ref{equ.P^c}), the average throughput of the network is
\begin{equation}
\label{equ.bar_R}
\begin{split}
\bar{R} &= P^c\bar{R_1} + (1-P^c)\bar{R_0}\\
& = WB(P^c\eta \bar{R_i^c}+(1-P^c\eta)\bar{R_i^n}).
\end{split}	
\end{equation}

\vspace{2mm}\begin{proposition}
	\label{p:1}
	The average data rate per unit bandwidth per second of the N-Coop link in the $i$th cluster is
	\begin{equation}
		\label{equ.bar_R_i^n3}
		\begin{split}
			\bar{R_i^n} = \log_2(Q_1(\alpha))-\log_2(Q_2(\alpha))-3,
		\end{split}
	\end{equation}
	where $Q_1(\alpha)\triangleq\int_{0}^{\sqrt{2}}r^{-\alpha}g(r)dr +8\int_{0}^{\sqrt{5}}r^{-\alpha}f(r)dr$,
	$Q_2(\alpha)\triangleq\int_{0}^{\sqrt{5}}r^{-\alpha}f(r)dr$, and $f(r)$ and $g(r)$ are in closed-form expression defined in Appendix \ref{a:1}.
\end{proposition}
\begin{IEEEproof}
	See Appendix \ref{a:1}.
\end{IEEEproof}\vspace{2mm}

In Proposition \ref{p:1}, $Q_1(\alpha)$ and $Q_2(\alpha)$ are easy to be computed numerically.
We can see that $\bar{R_i^n}$ only depends on the path loss exponent $\alpha$.

\vspace{2mm}\begin{proposition}
	\label{p:2}
	The average data rate per unit bandwidth per second of the Coop link in the $i$th cluster is
	\begin{equation}
		\label{equ.bar_R_i^c2}
		\begin{split}
			\bar{R_i^c} =\log_2(1+\frac{PD^{-\alpha}}{B\sigma^2}Q_1(\alpha)).
		\end{split}
	\end{equation}
\end{proposition}
\begin{IEEEproof}
	See Appendix \ref{a:2}.
\end{IEEEproof}\vspace{2mm}

By substituting  (\ref{equ.bar_R_i^n3}) and (\ref{equ.bar_R_i^c2}) into (\ref{equ.bar_R}), the average network throughput can be obtained as
\begin{equation}
	\label{equ.bar_R2}
	\begin{split}
		\bar{R} &= WBP^c\eta ( \log_2(Q_1(\alpha))-\log_2(Q_2(\alpha))-3)\\ &+WB(1-P^c\eta)\log_2(1+\frac{PD^{-\alpha}}{B\sigma^2}Q_1(\alpha).
	\end{split}
\end{equation}
\subsubsection{Average User Data Rate}

Since only one N-Coop user and one Coop user (if exists) are active in a cluster each time, with
round robin scheduling, the average data rates of N-Coop and Coop users can be
respectively obtained  from (\ref{equ.bar_R_i^n3}) and (\ref{equ.bar_R_i^c2}) as follows
\begin{align}
	\bar{R^n_u}   & =W(1-\eta) \mathbb{E}\{\frac{B R_i^n}{N^n} \}\nonumber \overset{(a)}{\approx} \frac{WB(1-\eta)\bar{R_i^n}}{\bar{N^n}} \nonumber \\&= \frac{WB(1-\eta)}{\bar{N^n}}\log_2(1+\frac{PD^{-\alpha}}{B\sigma^2}Q_1(\alpha)),  \nonumber\\
	\bar{R^c_u}  & =W\eta \mathbb{E}\{\frac{B R_i^c}{N^c} \}\nonumber  \overset{(b)}{\approx} \frac{WB\eta \bar{R_i^c}}{\bar{N^c}} \nonumber \\& = \frac{WB\eta(\log_2(Q_1(\alpha))-\log_2(Q_2(\alpha))-3) }{\bar{N^c}},\label{equ.bar_R^c_n}
\end{align}
where (a) and (b) come from the fact that $R_i^n$ and $N^n$ are independent random variables and the same to $R_i^c$ and $N^c$, thus $\mathbb{E}\{R_i^n/N^n\} = \mathbb{E}\{R_i^n\}\mathbb{E}\{1/N^n\} \approx
\mathbb{E}\{R_i^n\}/\mathbb{E}\{N^n\} = \bar{R_i^n}/\bar{N^n}$ according to (\ref{equ.1_approx}) and $\mathbb{E}\{R_i^c/N^c\} \approx \bar{R_i^c}/\bar{N^c}$ analogically. $\bar{N^c}=\mathbb{E}\{N^c\}$ and $\bar{N^n}=\mathbb{E}\{N^n\}$ are the average numbers of Coop users and N-Coop users,  respectively.

\vspace{2mm}\begin{proposition}
	\label{p:3}
	The average number of Coop users is
	\begin{equation}
		\label{equ.bar_N_c}
		\begin{split}
			\bar{N^c}
			&= \sum_{\Phi_\mathcal{N}}\prod_{i=1}^{B}\frac{K!\prod_{k=1}^{K_0} (P_k)^{n_{ik}}}{\prod_{j=1}^{K_0}n_{ij}!} \sum_{k=1}^{K}\sum_{i=1}^{B}\zeta(k) n_{ik},
		\end{split}
	\end{equation}
which can be approximated as
	\begin{equation}
		\label{equ.bar_N_c2}
		\begin{split}
			&\bar{N^c} \approx \bar{N^c_1} +\bar{N^c_2},
		\end{split}
	\end{equation}
	where $\zeta(k)$, $n_{ik}$, $\bar{N^c_1}$ and $\bar{N^c_1}$ are defined in Appendix \ref{a:3}.
\end{proposition}
\begin{IEEEproof}
	See Appendix \ref{a:3}.
\end{IEEEproof}\vspace{2mm}

Though we can use similar way to derive the average number of N-Coop users $\bar{N}^n$ as for  $\bar{N}^c$, the resulting expression is complicated.
Considering that $N^n+N^c+N^b=M$, we can obtain $\bar{N}^n$ by deriving the average number of cellular users $\bar{N^b}=\mathbb{E}\{N^b\}$.
Since all
requests follow a Zipf distribution independently, the number of users that can not fetch files via D2D is a random variable following a Binomial distribution and $N^b \sim B(M,1-\sum_{k=1}^{K}P_k)$. Therefore, $\bar{N^b}=M(1-\sum_{k=1}^{K}P_k)$. Then, the average number of N-Coop users is
\begin{equation}
	\label{equ.bar_N_n}
	\bar{N^n} = M - \bar{N^c} - \bar{N^b} .
\end{equation}

With the average number of Coop and N-Coop users $\bar{N^c}$ and $\bar{N^n}$, we can obtain the corresponding average user data rate  $\bar{R}^c_u$ and $\bar{R}^n_u$ using \eqref{equ.bar_R^c_n}.

\subsubsection{Joint Optimization of $\eta$ and $K$}
The bandwidth allocation factor and cluster size that maximize the average network throughput under the constraint of average user data rate can be optimized from the following problem
\begin{equation}
	\label{equ.opt2}
	\begin{aligned}
		&\max_{\eta,K}\,\, &&\bar{R} \\
		&s.t.\quad
		&&\bar{R}^c_u\geq \mu, ~~ \bar{R}^n_u \geq \mu,\\
		& && 0< \eta \leq 1, ~~ KB=M.
	\end{aligned}
\end{equation}

Since the number of users per cluster $K$ is an integer, we can find the joint optimal solution by first finding optimal $\eta$ for any given $K$ and then enumerating $K$ until the value of $\bar{R}$ computed by \eqref{equ.bar_R2} achieves the maximum under the two constraints.

By taking the derivative of $\bar{R}$  in (\ref{equ.bar_R}) with respect to $\eta$, we have $\frac{\partial \bar{R}}{\partial \bar{\eta}} = WBP^c(\bar{R_i^c}-\bar{R_i^n})$. The constraints $\bar{R}^c_u\geq \mu$ and $\bar{R}^n_u \geq \mu$ can be respectively rewritten as $\eta \leq \frac{WB\bar{R^n_i}-\mu\bar{N^n}}{WB\bar{R^n_i}}$ and $\eta \geq \frac{\bar{N^c}\mu}{WB\bar{R^c_i}}$ according to \eqref{equ.bar_R^c_n}. Therefore, if $\bar{R^c_i} \geq \bar{R^n_i}$, $\bar{R}$ is an increasing function of $\eta$, then the optimal solution of problem \eqref{equ.opt2} for any given $K$ is $\eta^*_K=\frac{WB\bar{R^n_i}-\mu\bar{N^n}}{WB\bar{R^n_i}}$; otherwise, $\bar{R}$ is a decreasing function of $\eta$, and $\eta^*_K=  \frac{\bar{N^c}\mu}{WB\bar{R^c_i}}$.

\vspace{2mm}\begin{proposition} \label{p:4}
	If $\bar{I_i} \geq B\sigma^2$, then $\bar{R^c_i} \geq \bar{R^n_i}$.
\end{proposition}

\begin{IEEEproof}
	See Appendix \ref{a:4}.
\end{IEEEproof}\vspace{2mm}

Since the power of interference among N-Coop users $\bar{I_i}$ defined in \eqref{equ.SINR_R^n} is much larger than noise variance $\sigma^2$ in D2D communications and $B$ is finite, the condition $\bar{I_i}\geq B\sigma^2$ in Proposition
\ref{p:4} is easy to be satisfied. Consequently, the optimal value of $\eta$ for a given $K$ is
\begin{equation}
	\label{equ.opt_eat}
	\eta^*_K = 1 - \frac{\mu\bar{N^n}}{WB\bar{R^n_i}},
\end{equation}
which only depends on the average data rate of N-Coop link ${\bar R^n_i}$ and the number of N-Coop users $\bar{N^n}$.


The joint optimal solution $K$ and $\eta$ can be found by one-dimensional searching and $\eta^*$ is with closed-form expression, and is hence with low complexity. Because $K^*$ and $\eta^*$ depend on $W$, $\mu$, $B$, $M$, $N_f$, $\beta$ and $N$, which are usually fixed for a long time, they are unnecessary to be updated frequently.

\section{Simulation and Numerical Results}

\label{sec:simulation} In this section, we validate the analysis and evaluate the performance of the proposed opportunistic cooperation strategy by simulation and numerical results.

In the simulation, we consider a square hotspot area with the side length $D_c=100$ m, where $M=180$ users are randomly located. Such a setting reflects relatively high user density as in \cite{3GPP.density}, where more than one user is located within an area of $10\times10$ $m^2$.
The
path-loss model is $37.6+36.8\log_{10}(r)$ \cite{JMY.JSAC}. Each user is with transmit power $P=23$ dBm. $W=20$ MHz, and $\sigma^2=-100$ dBm.
The file catalog size $N^f=300$, and each user is willing to cache $N=10$ files. The parameter of Zipf distribution is $\beta =0 \sim 1$. The user data rate constraint is
$\mu = 1 \sim 2$ Mbps.
This setup is used in the sequel unless otherwise specified.

\subsection{Impact of Cluster Size and Number of Total Users}
%

\begin{figure}[!htb]
	\centering
	\includegraphics[width=0.5\textwidth]{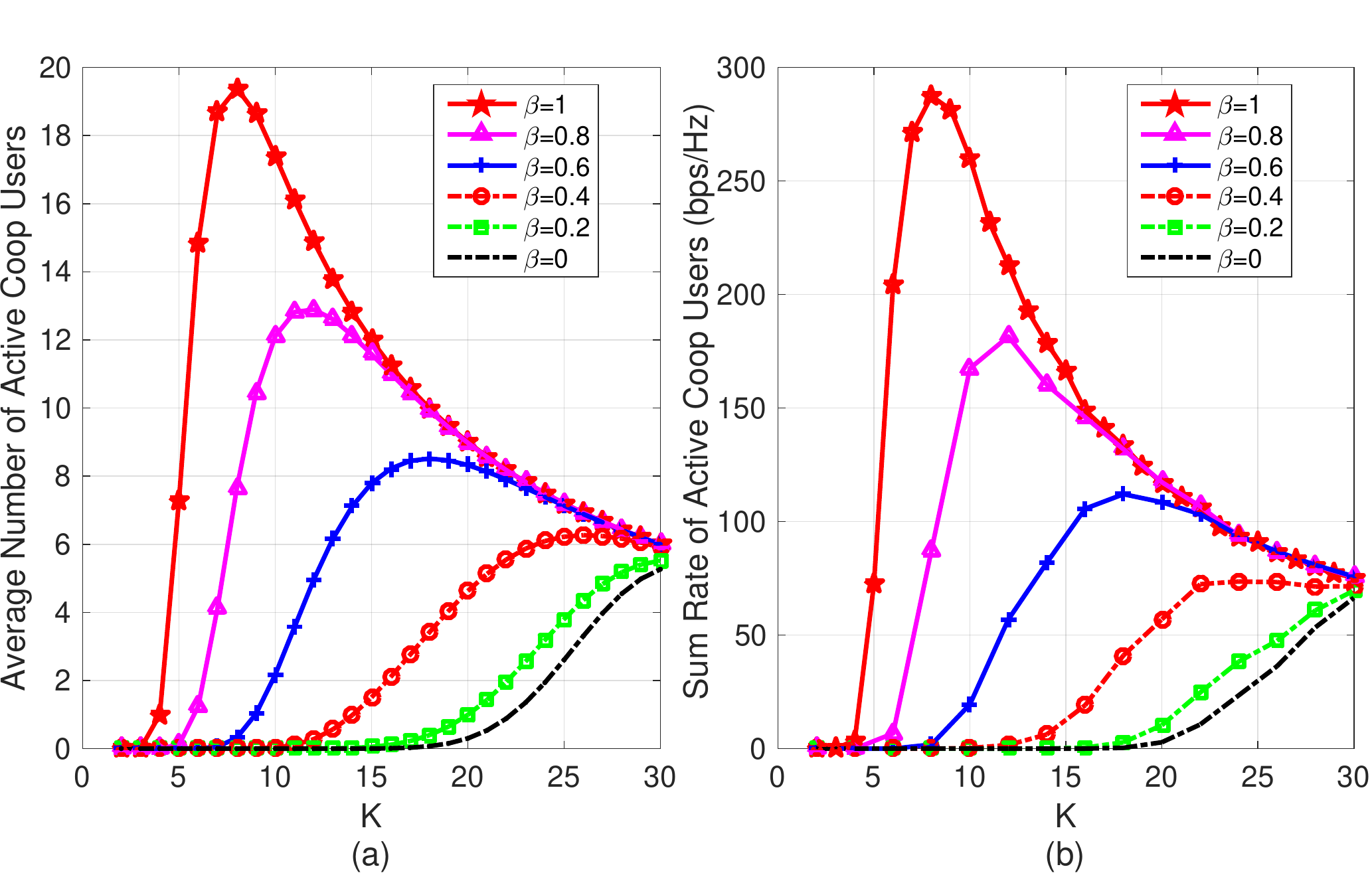}\\
	\caption{$\bar{N^a}$ and average sum rate of active Coop users versus $K$.}\label{fig.2}
\end{figure}

In Fig. \ref{fig.2}, we provide numerical results of the average number of active Coop users $\bar{N^a}$ obtained from \eqref{equ.E_N^a} and simulation results for the average sum rate of the active Coop users versus the number of users per cluster $K$. It is shown from Fig. \ref{fig.2}(a) that $\bar{N^a}$ increases with $\beta$. This is because with large $\beta$, a few popular files are requested by majority of users, which leads to high cooperative probability $P^c$. With the increase of $K$, $\bar{N^a}$ first increases and then decreases. It is shown from Fig. \ref{fig.2}(b) that with the growth of $K$, the average sum rate of active Coop users exhibits the same trends with $\bar{N^a}$, which agrees to the analysis in section III.A.

\begin{figure}[!b]
	\centering
	\includegraphics[width=0.5\textwidth]{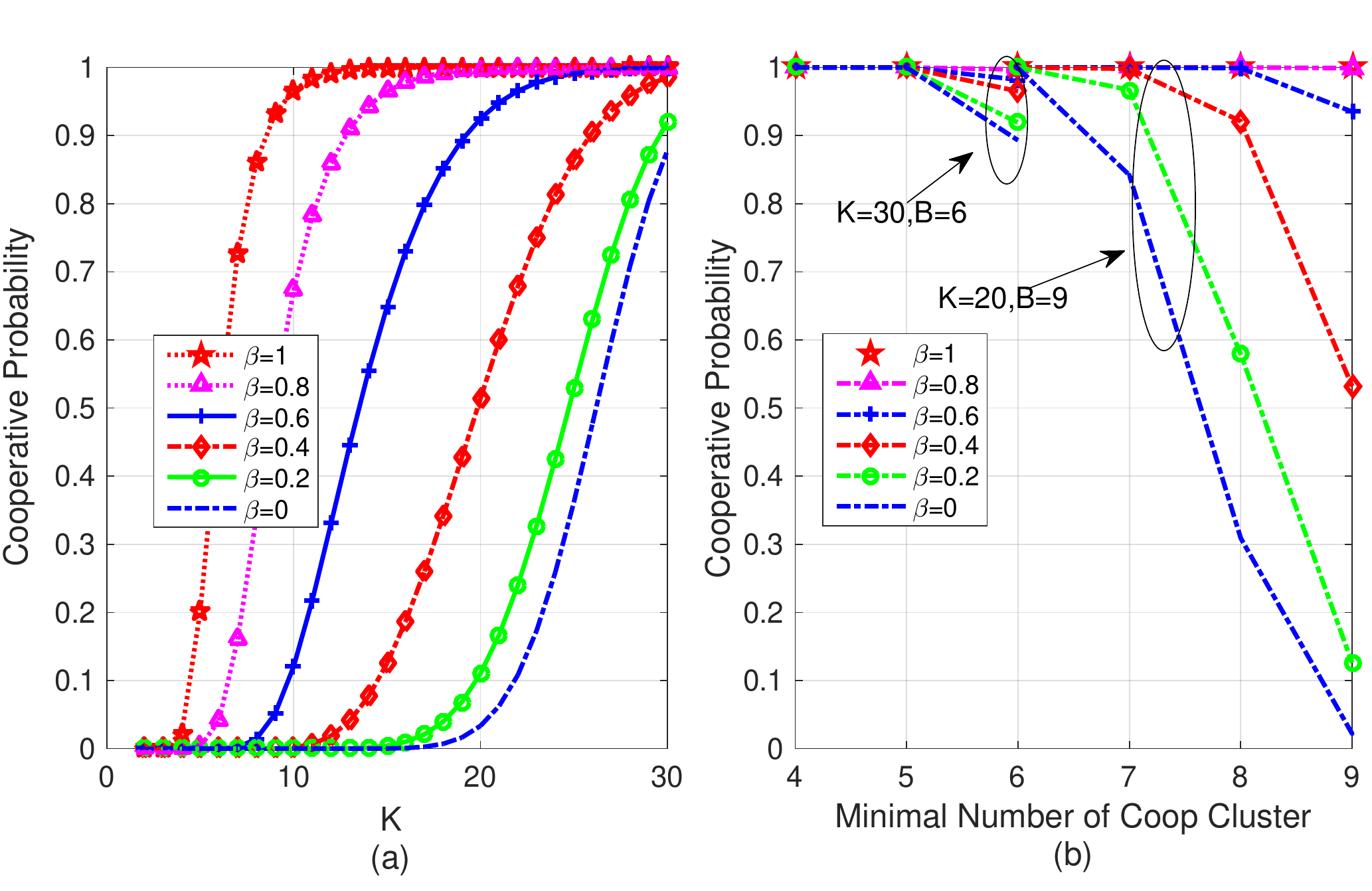}\\
	\caption{$P^c$ of (a) full cooperative, and (b) partial cooperation.}\label{fig.3}
	
\end{figure}

In Fig. \ref{fig.3}, we simulate the cooperative probability $P^c$. It is shown from Fig. \ref{fig.3}(a) that $P^c$
increases with $K$, which agrees with the analysis after (\ref{equ.P^c}). Moreover, the cooperative probability is high although  the full cooperation is allowed only when all clusters hit the same file group, especially when $\beta$ is not small, say $\beta > 0.4$.
In Fig. \ref{fig.3}(b), we show the impact of partial cooperation by changing the minimal number of clusters allowed to cooperate.\footnote{We set such a minimal number because when we allow fewer clusters to cooperate, the multiplexing gain will reduce despite that $P_c$ will be higher, and then the resulting throughput will reduce.  When this minimal number of clusters is six, it becomes the full cooperation strategy.} As expected, $P^c$ can be improved if we allow partial cooperation among clusters, but not significant. Therefore, the throughput gain from partial cooperation over full cooperation is marginal as to be illustrated later.

\begin{figure}[!htb]
	\centering
	\includegraphics[width=0.5\textwidth]{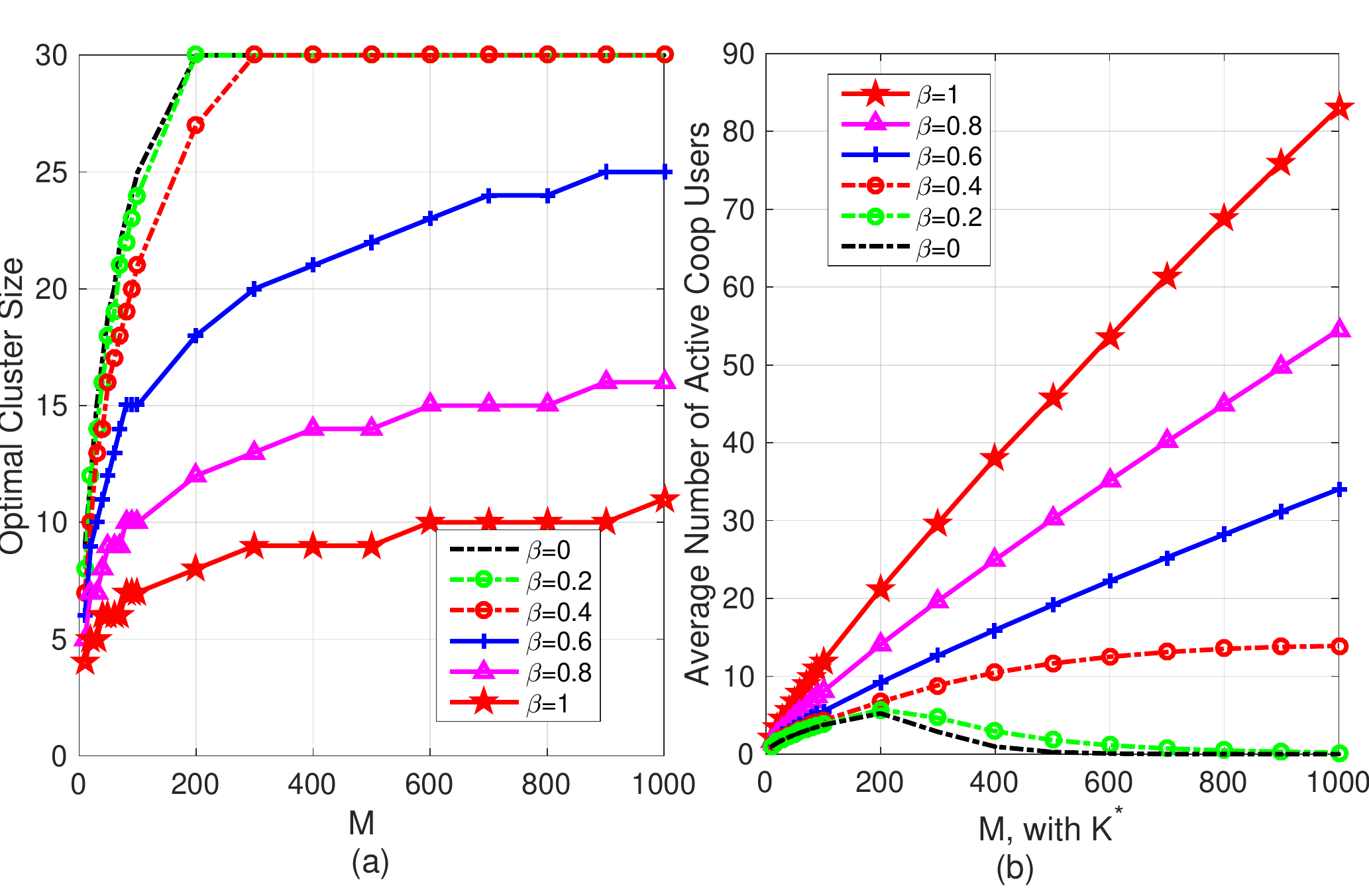}\\
	\caption{$K^*$ and $\bar{N^a}$ versus the number of total users in the hotpot $M$.}\label{fig.1a}
\end{figure}

In Fig. \ref{fig.1a}, we provide numerical results of the optimal cluster size $K^*$ and the resulting average number of active Coop users $\bar{N^a}$ for different number of total users in the hotspot area $M$. When $M=1000$, there are $10$ users in an area of $10 \times10$ $m^2$, corresponding to a very high traffic load \cite{3GPP.density}. It is shown from Fig. \ref{fig.1a}(a) that $K^*$ decreases with $\beta$ as expected. With the growth of $M$, $K^*$ first increases  and then approaches a constant that equals to $K_0=N^f/N$. This is because when $K=K_0$, all files in the catalog can be cached at the users in each cluster. Assigning more than $K_0$ users to each cluster can not increase the number of Coop users. It is shown from Fig. \ref{fig.1a}(b) that  with the growth of $M$, the average number of active Coop users monotonously increases when $\beta$ is large but first increases and then decreases when $\beta$ is small. This implies that if the user density is large but $\beta$ is small, the proposed opportunistic cooperation strategy may be not useful.

\subsection{Accuracy of the Approximations}
\begin{figure}[!htb]
	\centering
	\includegraphics[width=0.5\textwidth]{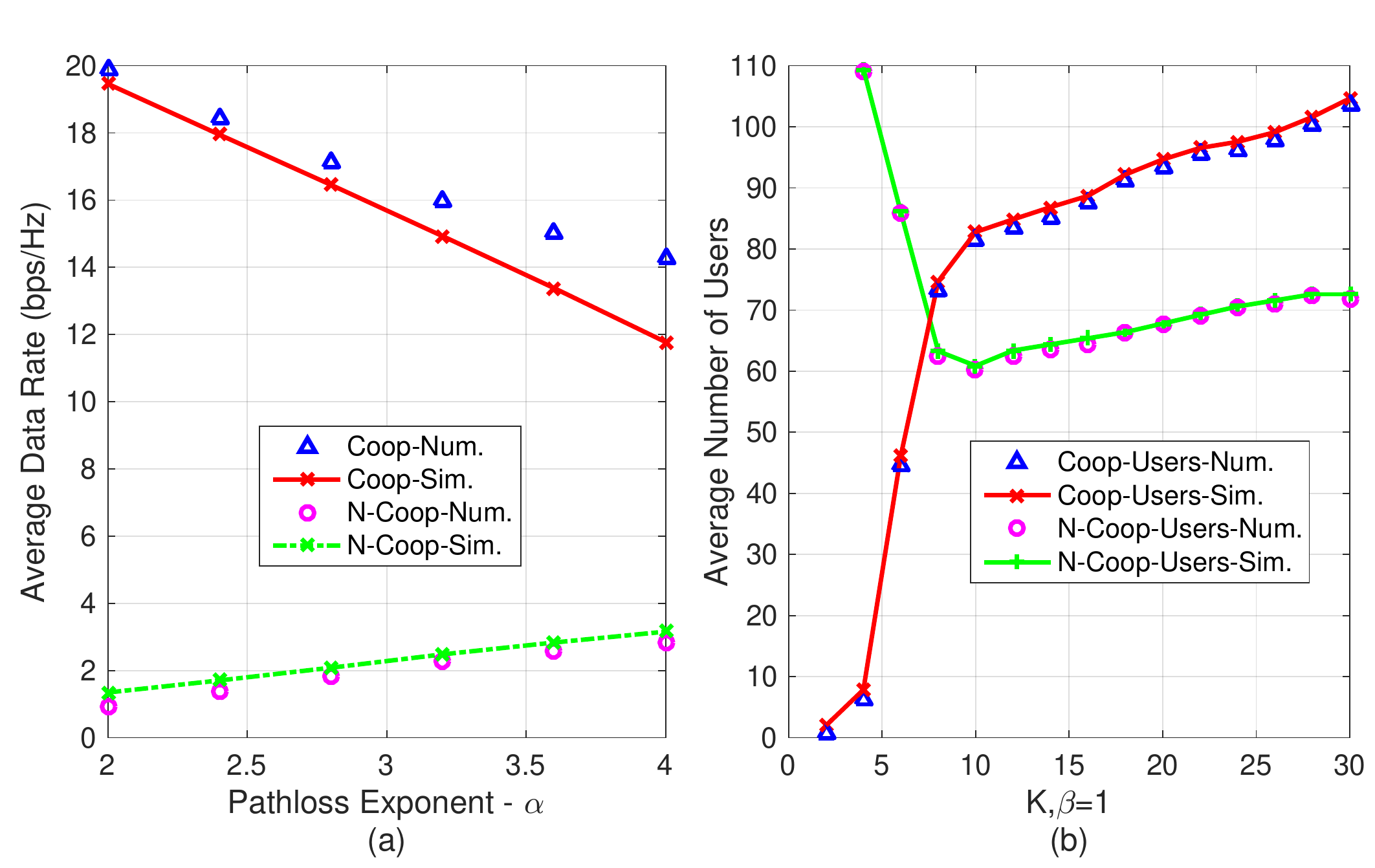}\\
	\caption{Average data rate
per unit bandwidth per second and number of users.  ``Num.'': Numerical results, ``Sim.'': Simulation results.}\label{fig.4}
\end{figure}

In Fig. \ref{fig.4}, we evaluate the accuracy of all the approximations used in deriving the throughput and the number of users by simulation and numerical results.
In Fig.
\ref{fig.4}(a), the numerical results of $\bar{R^c_i}$ and $\bar{R^n_i}$ are respectively obtained from (\ref{equ.bar_R_i^n3}) and (\ref{equ.bar_R_i^c2}) by changing the path loss exponent $\alpha$ from $2$ to $4$. We can see that with the growth of $\alpha$, the average data rate per unit bandwidth per
second of the Coop link decreases but that for the N-Coop link increases. This is because both signal and interference power decrease when $\alpha$ increases, but the interference power decreases more rapidly due to larger distance of interference link than that of signal link.
In Fig. \ref{fig.4}(b), the numerical results of the number of Coop users $\bar{N^c}$ and the number of N-Coop users $\bar{N^n}$ are  respectively obtained from \eqref{equ.bar_N_c2} and \eqref{equ.bar_N_n}. As expected, when $K$ increases, the number of Coop users increases due to the decrease of the number of clusters. However, the number of N-Coop users first decreases and then increases slowly with $K$. This is because with the growth of $K$, more files can be cached in each cluster and thus the total number of D2D users increases, and the number of N-Coop users changes as in \eqref{equ.bar_N_n}.
We can see that the approximations are accurate, except the average data rate
per unit bandwidth per second of the Coop link when $\alpha$ is large, which comes from the first order approximation in \eqref{equ.bar_R_i^n2}.

\subsection{Optimal Bandwidth Allocation and Network Throughput}

\begin{figure}[!htb]
	\centering
	\includegraphics[width=0.5\textwidth]{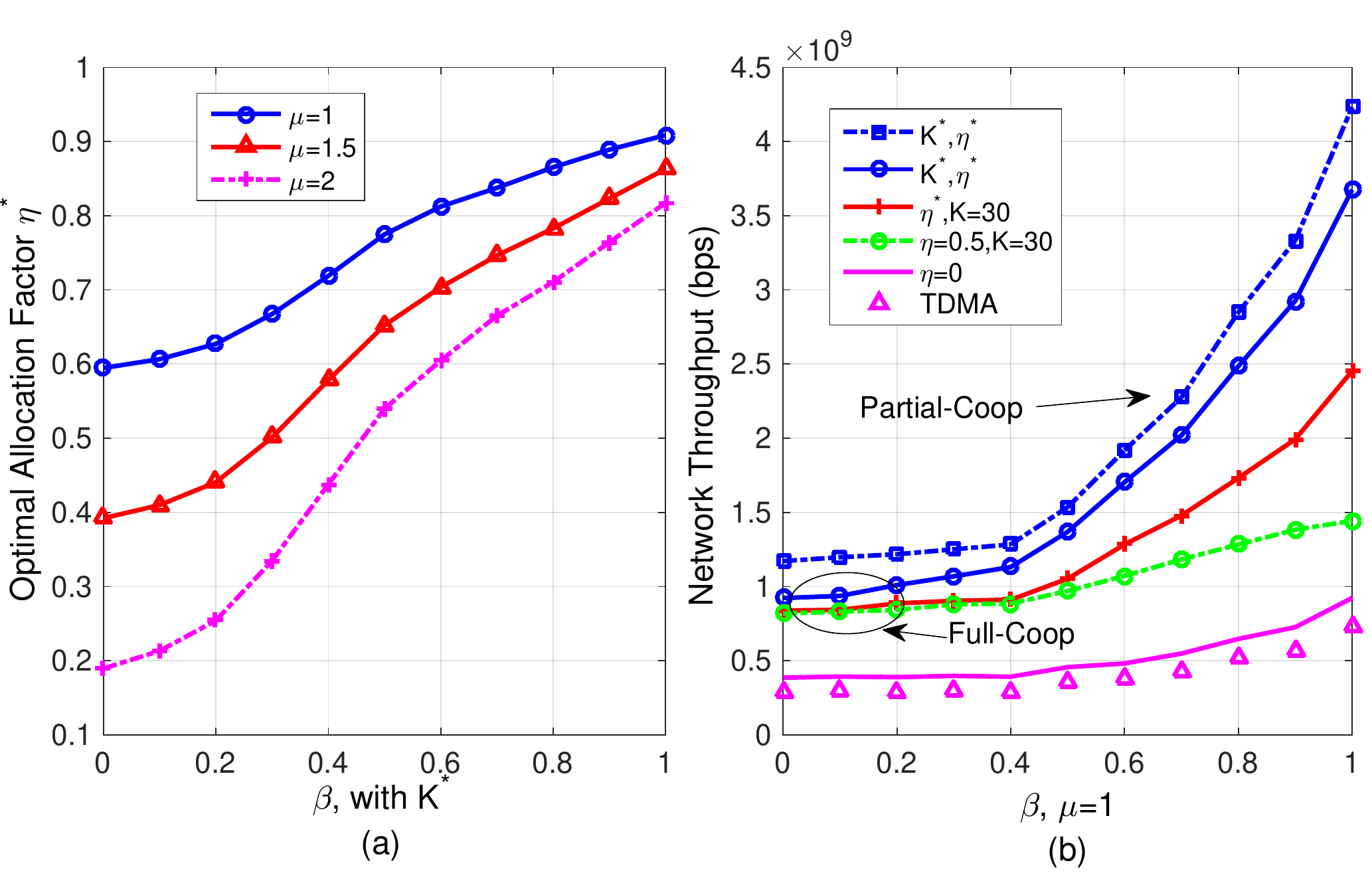}\\
	\caption{$\eta^*$ and maximal average network throughput versus $\beta$. }\label{fig.5}
\end{figure}

In Fig. \ref{fig.5}(a), we present the optimal solution of problem  (\ref{equ.opt2}) $\eta^*$ versus the Zipf parameter $\beta$. We can see that $\eta^*$ increases with $\beta$. This is because the number of interference-free links, $\bar{N^a}$, increases with $\beta$, and then allocating more bandwidth to Coop users can increase the network throughput. $\eta^*$
decreases as $\mu$ increases, because more bandwidth is needed for N-Coop users to support higher user data rate.

In Fig. \ref{fig.5}(b), we provide simulation results for the maximal network throughput. In the legends, ``$\eta=0$" refers to the strategy in  \cite{Golrezaei.TWC} (without interference coordination) and ``TDMA" is the strategy in \cite{JMY.JSAC} (without strong inter-cluster
interference), which serve as the baselines for comparison.  ``$\eta=0.5, K=30$" refers a cooperation strategy without optimizing $K$ and $\eta$, which allocates equal bandwidth to Coop users and N-Coop users. ``$\eta^*, K=30$" refers to a cooperation strategy without optimizing $K$ but only optimizing $\eta$. We can see that optimizing the bandwidth allocation becomes necessary when $\beta >0.5$, while optimizing the cluster size is always necessary but the gain from optimization grows with $\beta$. With $K^*$ and $\eta^*$, the throughput gain over the baseline for $\beta=1$ is  $400\% \sim 500\%$, which demonstrates that the proposed opportunistic cooperation strategy can boost the network throughput remarkably. Even when $\beta=0$, where file popularity follows a uniform distribution, the throughput gain is still $60\% \sim 80\%$. We also demonstrate the performance of ``Partial-Coop", i.e., allowing cooperation among less than $B$ clusters. It is shown that the performance can be improved by allowing partial cooperation, but the gain is marginal.

\section{Conclusions}
\label{sec:conclusion} In this paper, we proposed an opportunistic cooperation strategy for cache-enabled D2D communications. We jointly optimized the cluster size and the bandwidth allocated to Coop and N-Coop users to maximize the network throughput with minimal user data rate constraint. Simulation results showed that the proposed strategy can boost the throughput even when the content popularity follows a uniform distribution, and the gain over existing strategies is remarkable when the popularity distribution is more skewed.

\appendices
\numberwithin{equation}{section}

\section{Proof of Proposition \ref{p:1}}
\label{a:1}
Without cooperation, the closest DT in the same cluster of its corresponding DR delivers the requested file to the DR, which treats the inter-cluster interference as noise when decoding the desired signal. The signal to interference plus noise ratio (SINR) at the DR of the active N-Coop link in the $i$th cluster can be expressed as
\begin{equation}
	\label{equ.SINR_R^n}
	\gamma_i^n = \frac{P|h_{ii}|^2r_{ii}^{-\alpha}}{I_i+\sigma^2},
\end{equation}
where $P$ is the transmit power, $\sigma^2$ is the variance of white Gaussian noise, $I_i = P\sum_{j=1,j \neq i}^{B}r_{ij}^{-\alpha}|h_{ij}|^2$
is the total power of inter-cluster interference, $h_{ij}$ and $r_{ij}$ are respectively the channel coefficient and distance between the DT  and the DR, $\alpha$ is the path loss exponent, and both the interference channel coefficient $h_{ij}$ $(i\neq j)$ and the desired channel coefficient $h_{ii}$ follow a complex Gaussian distribution with zero mean and unit variance.

Due to the short distance between D2D links, the D2D network is interference-limited and hence the noise can be ignored, i.e., $I_i \gg \sigma^2$.

Then,  from \eqref{equ.SINR_R^n}
the data rate per unit bandwidth per second for the N-Coop link in the $i$th cluster is $R_i^n = \log_2(1+\frac{P|h_{ii}|^2r_{ii}^{-\alpha}}{I_i})$.

Considering that $|h_{ij}|^2$ follows an Exponential distribution, which is a special case of the Gamma
distribution, the interference power $I_i$ (which is a sum of random variables following a Gamma distribution)  can be approximated as a Gamma distribution
\cite{Gamma.2011}. Further consider that for a Gamma distributed random variable $X$ with parameters $k$ and $\theta$,
$\mathbb{E}\{\ln(X)\}=\psi(k)+\ln(\theta)$, where $\psi(k)$ is the Digamma function \cite{abramowitz1964}. Then, the average  data rate per unit bandwidth per second  of the N-Coop link can be obtained according to Proposition 9 in \cite{Gamma.2011} as
\begin{equation}
	\label{equ.E_h_R_i^n}
	\mathbb{E}_{\mathbf{h}}\{R_i^n\} \approx \log_2(1+\frac{Pr_{ii}^{-\alpha}}{\bar{I_i}}),
\end{equation}
where $\mathbb{E}_{\mathbf{h}}\{\cdot\}$ represents the expectation taken over small scale channel fading, $\bar{I_i} = P\sum_{j=1,j \neq i}^{B}r_{ij}^{-\alpha} $ is the average total power of the inter-cluster interference.

Since channel fading and user location are distributed independently, the average data rate per unit bandwidth per second  of the N-Coop link taken over both channel fading and user location
can be obtained as
\begin{equation}
	\label{equ.bar_R_i^n}
	\begin{split}
		&\bar{R_i^n} =\mathbb{E}_{\textbf{p}} \{ \log_2(1+\frac{Pr_{ii}^{-\alpha}}{\bar{I_i}}) \},
	\end{split}
\end{equation}
where $\mathbb{E}_{\mathbf{p}}\{\cdot\}$ denotes the expectation taken over user location.

Because the joint probability density function (pdf) of the distances among D2D users is hard to obtain, we introduce the first order approximation to derive
the expression of $\bar{R_i^n}$. Specifically, for a random variable $X$, the expectation of a function of $X$, $\varphi( X)$, can be approximated as
\cite{approx}
\begin{equation}
	\label{equ.1_approx}
	\begin{split}
		\mathbb{E}\{\varphi( X)\}& = \mathbb{E}\{\varphi(\mu_x +  X - \mu_x )\}  \\
		&\approx \mathbb{E}\{\varphi(\mu_x) + \varphi'(\mu_x)( X-\mu_x)\} \\
		&= \varphi(\mu_x),
	\end{split}
\end{equation}
where $\mu_x = \mathbb{E}\{ X\}$ and the approximation is accurate when the variance of $X$ is small. With this approximation, $\bar{R_i^n}$ in (\ref{equ.bar_R_i^n}) can be approximated as
\begin{equation}
	\label{equ.bar_R_i^n2}
	\bar{R_i^n} \approx \log_2(\mathbb{E}_{\textbf{p}}\{Pr_{ii}^{-\alpha}+\bar{I_i}\}) - \log_2(\mathbb{E}_{\textbf{p}}\{\bar{I_i}\}).
\end{equation}
The pdf of the signal link distance $r_{ii}$ can be obtained from \cite{PDF}
by variable substitution $r=r_{ii}/D$ as
\begin{eqnarray}
	\label{equ.g_r_ii}
	g(r)=\frac{1}{D}
	\begin{cases}
		2r(r^2-4r+\pi), &0\leq r<1 \\
		8r\epsilon-2r(r^2+2) \\+4r(\arcsin(\frac{1}{r})\\-\arccos(\frac{1}{r})), &1\leq r<\sqrt{2}\\
	\end{cases}.
\end{eqnarray}

To simplify the analysis, the interference link distance $r_{ij}$ is assumed to have the same distribution $f(r)$, where $r=r_{ij}/D$. We can derive the pdf of the interference link distance $r_{ij}$ as follows.
\begin{figure}[!t]
	\centering
	\includegraphics[width=0.3\textwidth]{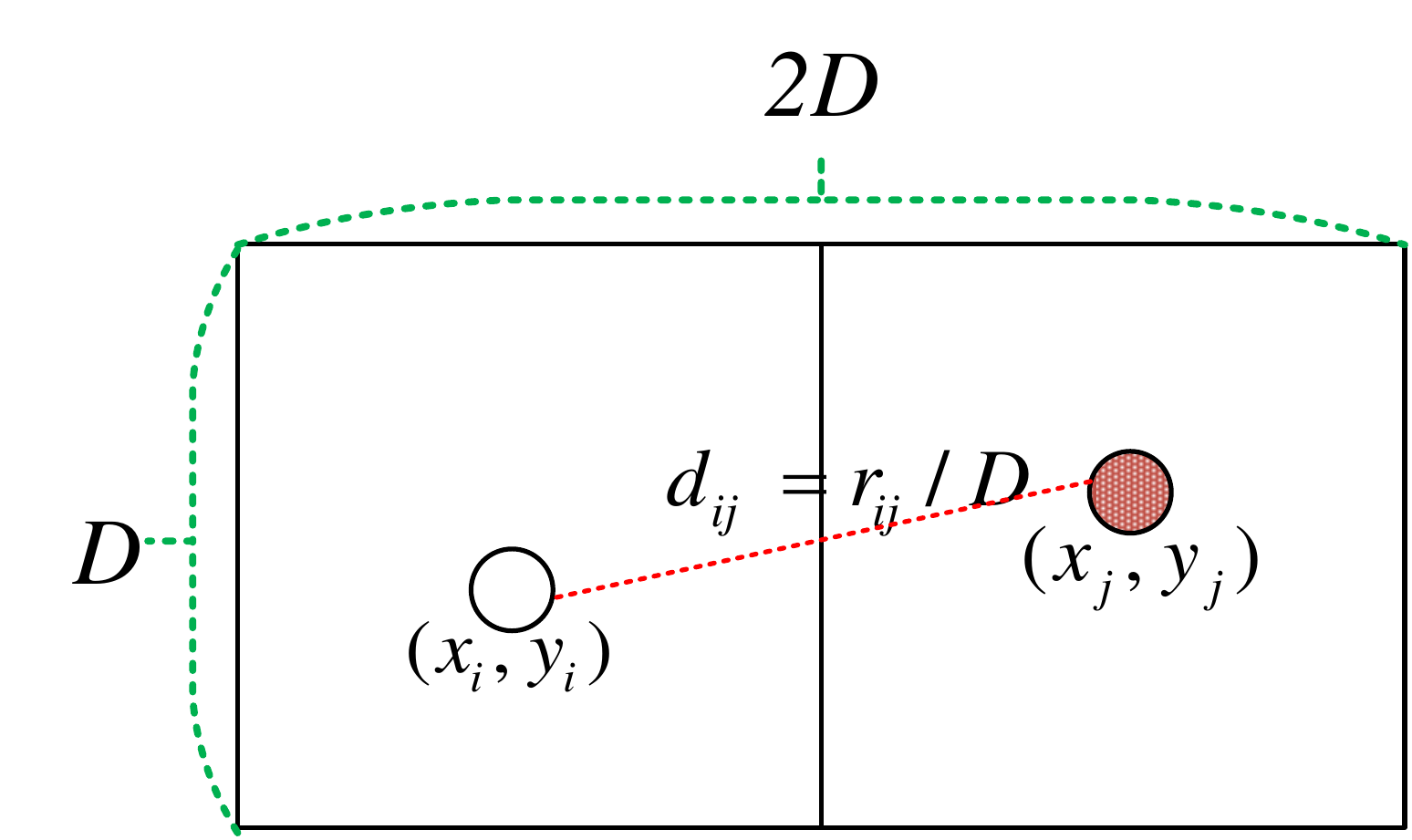}\\
	\caption{The distance of users in two adjacent cluster}\label{fig.6}
\end{figure}	

Denote the position of the DR in the $i$th cluster as $(x_i,y_i)$ and the position of the DT in the $j$th cluster as $(x_j,y_j)$, as illustrated in Fig. \ref{fig.6}. The link distance between them normalized by the cluster side $D$ can be expressed as $d_{ij} = \frac{\sqrt{(x_i-x_j)^2 + (y_i-y_j)^2}}{D} =\sqrt{(\Delta x)^2 + (\Delta y)^2}$, where $\Delta x = \frac{x_i-x_j}{D}$ and $\Delta y = \frac{y_i-y_j}{D}$. The pdf of $|\Delta y|$ can be obtained according to \cite{PDF} as
\begin{eqnarray}
	\label{equ.dy.pdf}
	p_{\Delta y} (v)=
	\begin{cases}
		{2(1-v)}, &0 \leq v \leq 1 \\
		0, &\text{otherwise}\\
	\end{cases}.
\end{eqnarray}
Analogically, the pdf of $|\Delta x|$ can be obtained as
\begin{eqnarray}
	\label{equ.dx.pdf}
	p_{\Delta x} (u)=
	\begin{cases}
		1-|1-u|, &0 \leq u \leq 2 \\
		0, &\text{otherwise}\\
	\end{cases}.
\end{eqnarray}

Then, the cumulative probability distribution function (cdf) of $d_{ij}$ is
\begin{equation}
	\label{equ.r.cdf}
	\begin{split}
		F_d(r) &= \mathbb{P} \{d_{ij} \leq r\} =  \mathbb{P}  \{\sqrt{|\Delta x|^2 + |\Delta y|^2} \leq r \} \\
		&= \iint du dv p_{\Delta x,\Delta y}(u,v) \\
		&\stackrel{(a)}{=} \iint du dv p_{\Delta x} (u) p_{\Delta y} (v),
	\end{split}
\end{equation}
where (a) comes from the fact that $|\Delta x|$ and $|\Delta y|$ are independent random variables.

When $0 \leq r \leq 1$, from \eqref{equ.dx.pdf} and \eqref{equ.dy.pdf}, we can obtain $p_{\Delta x,\Delta y}(u,v) = 2u(1-v)$, where $0 \leq u,v \leq 1$. Then, the cdf in  \eqref{equ.r.cdf} can be derived as
\begin{equation}
	\label{equ.r.cdf.1}
	\begin{split}
		F_d(r) &=  \iint du dv p_{\Delta x,\Delta y}(u,v) \\
		&= \int_{0}^{r}dv \int_{0}^{\sqrt{r^2-v^2}}2u(1-v) du \\
		&= \frac{2}{3}r^3 - \frac{1}{4}r^4,
	\end{split}
\end{equation}

When $1 \leq r \leq \sqrt{2}$, we have
\begin{eqnarray}
	\nonumber
	\label{equ.dxdy.pdf2}
	p_{\Delta x,\Delta y}(u,v)=
	\begin{cases}
		2u(1-v), &0 \leq u \leq 1, 0 \leq v \leq 1 \\
		2(2-u)(1-v), &1 \leq u \leq 2, 0 \leq v \leq 1\\
		0, &\text{otherwise}\\
	\end{cases},
\end{eqnarray}
and the corresponding cdf can be derived as
\begin{equation}
	\label{equ.r.cdf.2}
	\begin{split}
		&F_d(r)= \\
		&\int_{0}^{\sqrt{r^2-1}}dv \left( \int_{0}^{1} 2u(1-v) du +\int_{1}^{\sqrt{r^2-v^2}}2(2-u)(1-v) du \right)\\
		&+\int_{\sqrt{r^2-1}}^{1}dv \int_{0}^{\sqrt{r^2-v^2}}2u(1-v) du\\
		&= \frac{5}{4} + \epsilon \left(  -2r^2+(1+r^2)\epsilon + \frac{2}{3}\epsilon^2 -\frac{1}{2}\epsilon^2   \right) \\
		& + r^2 \left( 2 \sin^{-1}(\frac{3}{r}) + \frac{1}{2} - \frac{4}{3}r \right) ,
	\end{split}
\end{equation}
where $\epsilon \triangleq \sqrt{r^2-1}$.

Analogically, when $\sqrt{2} \leq r \leq 2 $,  the cdf can be derived as
\begin{equation}
	\label{equ.r.cdf.3}
	\begin{split}
		F_d(r) &=  \int_{0}^{1}dv   \int_{0}^{1} 2u(1-v) du\\
		& +\int_{0}^{1}dv  \int_{1}^{\sqrt{r^2-v^2}}2(2-u)(1-v) du \\
		&= -\frac{11}{12} + 2\epsilon - \frac{r^2}{2} + 2r^2\sin^{-1}(\frac{1}{r}) + \frac{4}{3}(\epsilon^3-r^3).
	\end{split}
\end{equation}

When $ 2  \leq r \leq \sqrt{5} $, the cdf can be derived as
\begin{equation}
	\label{equ.r.cdf.4}
	\begin{split}
		F_d(r) &= 1- \int_{\sqrt{r^2-4}}^{1}dv\int_{\sqrt{r^2-v^2}}^{2} 2u(1-v) du \\
		&= -\frac{45}{12} - \frac{r^2}{2} - \frac{r^4}{2} -2r^2 \left(   \sin^{-1}(\frac{\xi}{r}) - \sin^{-1}(\frac{1}{r}) \right)\\
		& + r^2\xi + 2\epsilon -\frac{1}{3} \xi^3 + \frac{4}{3} \epsilon^3 +\frac{1}{4}\xi,
	\end{split}
\end{equation}
where  $\xi \triangleq \sqrt{r^2-4}$.

Finally, by combining \eqref{equ.r.cdf.1} \eqref{equ.r.cdf.2} \eqref{equ.r.cdf.3} and \eqref{equ.r.cdf.4}, and considering the pdf $f(r) = \frac{dF(r)}{dr}$, the pdf of the interference link distance $r_{ij}$ is
\begin{eqnarray}
	\label{equ.f_r_ij}
	f(r)=\frac{1}{D}
	\begin{cases}
		2r^2-r^3, &0\leq r<1 \\
		2r-4r^2+2r^3\\
		-2r\epsilon+\frac{2r}{\epsilon}\\
		-\frac{2r^3}{\epsilon}+4r\arcsin(\frac{\epsilon}{r}), &1\leq r<\sqrt{2}\\
		4r\epsilon+4r\arcsin(\frac{1}{r})\\
		-r-4r^2, &\sqrt{2}\leq r<2\\
		-5r-r^3+4r\epsilon\\
		-4r\arcsin(\frac{\xi}{r})-\arcsin(\frac{1}{r})\\
		-\frac{4r}{\xi}+r\xi+\frac{r^3}{\xi}, &2\leq r<\sqrt{5}\\
	\end{cases}
\end{eqnarray}
where $r = r_{ij}/D$, $\epsilon \triangleq \sqrt{r^2-1}$, and $\xi \triangleq \sqrt{r^2-4}$. 

Since the interference generated by DTs far away from the DR can be ignored due to pathloss, we only consider dominant interference generated from the nearest eight clusters around the $i$th cluster as shown in Fig. \ref{fig.ici-d2d}. Then, by substituting the pdf of interference and signal link distance into \eqref{equ.bar_R_i^n2}, we can obtain the average data rate per unit bandwidth per second of the N-Coop link in the $i$th cluster as
\begin{equation}
	\label{a:equ.bar_R_i^n3}
	\begin{split}
		\bar{R_i^n} &\approx \log_2\left(\int_{0}^{\sqrt{2}D}Pr_{ii}^{-\alpha}g(\frac{r_{ii}}{D})dr_{ii} \right.\\ &\left.+8\int_{0}^{\sqrt{5}D}Pr_{ij}^{-\alpha}f(\frac{r_{ij}}{D})dr_{ij}\right)\\
		& - \log_2\left(8\int_{0}^{\sqrt{5}D}Pr_{ij}^{-\alpha}f(\frac{r_{ij}}{D})dr_{ij} \right) \\
		&= \log_2\left(\int_{0}^{\sqrt{2}}Pr^{-\alpha}g(r)dr +8\int_{0}^{\sqrt{5}}Pr^{-\alpha}f(r)dr\right) \\
		&- \log_2\left(8\int_{0}^{\sqrt{5}}Pr^{-\alpha}f(r)dr\right) \\		
		&= \log_2(Q_1(\alpha))-\log_2(Q_2(\alpha))-3,
	\end{split}
\end{equation}
This proves Proposition \ref{p:1}.

\section{Proof of Proposition \ref{p:2}}
\label{a:2}
In \emph{Mode 1}, the cooperative DTs jointly  transmit the requested files to the Coop users with zero-forcing beamforming, which is of low complexity and hence practical. Then, the SINR at the DR of the active Coop link in the $i$th cluster can be expressed as
\begin{equation}
	\label{equ.SINR_R^c}
	\begin{split}
	\gamma_i^c &= \frac{P\| \textbf{h}_{i}\|^2\delta_i}{\sigma^2} \\
	&\approx  \frac{P\sum_{j=1}^{B}r_{ij}^{-\alpha}| h_{ij}|^2}{B\sigma^2} ,
	\end{split}
\end{equation}
where $\textbf{h}_{i}=[\sqrt{r_{i1}^{-\alpha}}h_{i1},\sqrt{r_{i2}^{-\alpha}}h_{i2},...,\sqrt{r_{iB}^{-\alpha}}h_{iB}]$ is the composite channel vector
between all DTs and the DR, $0\le \delta_i \le 1$, and a larger value of $\delta_i$ indicates a better orthogonality
between $\textbf{h}_{i}$ and $\textbf{h}_{j}$ for $i \ne j$. The approximation comes from the fact $\delta_i \approx (BN^t-B+1)/{B}={1}/{B}$ \cite{ZQ.TVT13}, where $N^t$ is the number of antennas per DT and $N^t=1$ in this paper. This approximation is accurate when the variance of $\delta_i$ is small.

Using the same approximation as deriving (\ref{equ.E_h_R_i^n}), the average data rate per unit bandwidth per second of the Coop link  is obtained as
\begin{equation}
	\label{equ.bar_R_i^c}
	\begin{split}
		&\bar{R_i^c} \approx \mathbb{E}_{\textbf{p}} \{ \log_2(1+\frac{P\sum_{j=1}^{B}r_{ij}^{-\alpha}}{B\sigma^2}) \}.
	\end{split}
\end{equation}

By applying the first-order approximation in (\ref{equ.1_approx}), using (\ref{equ.g_r_ii}) and (\ref{equ.f_r_ij}), and only considering dominant signal, we can obtain the average data rate per unit bandwidth per second for the Coop link as
\begin{equation}
\label{equ.bar_R_i^c2_app}
\begin{split}
\bar{R_i^c} &\approx \log_2\left(B\sigma^2 +8\int_{0}^{\sqrt{5}D}Pr_{ij}^{-\alpha}f(\frac{r_{ij}}{D})dr_{ij} \right.\\
&\left. + \int_{0}^{\sqrt{2}D}Pr_{ii}^{-\alpha}g(\frac{r_{ii}}{D})dr_{ii} \right) - \log_2\left( B \sigma^2 \right) \\
&=\log_2(1+\frac{PD^{-\alpha}}{B\sigma^2}Q_1(\alpha)).
\end{split}
\end{equation}
This proves Proposition \ref{p:2}.

\section{Proof of Proposition \ref{p:3}}
\label{a:3}
Denote the number of users in the $i$th cluster who request files in $\mathcal{G}_k$ as $n_{ik}$,  $1 \leq k \leq K_0$, $1
\leq i \leq B$.

Since the users request files independently, the probability that the combination of the numbers of users in the $i$th cluster who request files in each file group is $\{n_{i1},n_{i2},...,n_{i{K_0}}\} \triangleq \mathcal{N}_{i}$
can be derived as
\begin{equation}
	\label{equ.P_N_i}
	\begin{split}
		{p_{\mathcal{N}_{i}}}& = \prod_{m=1}^{K_0}C_{K-\sum_{j=1}^{m-1}n_{ij}}^{n_{im}}\prod_{k=1}^{K_0} (P_k)^{n_{ik}} \\
		&\overset{(a)}{=} \frac{K!\prod_{k=1}^{K_0} (P_k)^{n_{ik}}}{\prod_{j=1}^{K_0}n_{ij}!},
	\end{split}
\end{equation}
where $(a)$ comes from $C_{n}^{m}C_{n-m}^{k}=\frac{n!}{m!k!(n-m-k)!}$.

Only when all $B$ clusters hit a file group $\mathcal{G}_k$ and $k \leq K$ (i.e., $n_{ik}>0$  is satisfied for $k \leq K$  and any $i$, $1 \leq i \leq
B$), the users requesting the files within $\mathcal{G}_k$ are Coop users, and we call $\mathcal{G}_k$ a \emph{hit file group}.

The number of Coop users for
all hit file groups can be obtained as
\begin{equation}
	\label{equ.N_c}
	\begin{split}
		&{N^c} = \sum_{k=1}^{K}\sum_{i=1}^{B}\zeta(k) n_{ik},
	\end{split}
\end{equation}
where $\zeta(k)=\lceil \sum_{i=1}^{B} u(n_{ik})-B+1) \rceil ^+$ ($k\leq K$) indicates that whether $\mathcal{G}_k$ is a \emph{hit file group}, the function $u(x)=1$ when
$x>0$, otherwise $u(x)=0$, and $\lceil \Lambda \rceil ^+= \max(\Lambda,0)$.


Denote $\mathcal{N}=\{\mathcal{N}_1,\mathcal{N}_2,...,\mathcal{N}_B\}$, and
$\Phi_\mathcal{N}=\{\mathcal{N}|n_{ik}\geq 0,\sum_{k=1}^{K_0}n_{ik}=K\}$ represents a set of all possible combinations of $\mathcal{N}$. Then, by taking average of $N^c$ in
(\ref{equ.N_c}) over $\Phi_\mathcal{N}$, we can derive the average number of Coop users as
\begin{equation}
	\label{equ.bar_N_c_app}
	\begin{split}
		\bar{N^c}
		&=  \sum_{\Phi_\mathcal{N}}\prod_{i=1}^{B}{p_{\mathcal{N}_{i}}}{N^c}  \\
		&= \sum_{\Phi_\mathcal{N}}\prod_{i=1}^{B}\frac{K!\prod_{k=1}^{K_0} (P_k)^{n_{ik}}}{\prod_{j=1}^{K_0}n_{ij}!} \sum_{k=1}^{K}\sum_{i=1}^{B}\zeta(k) n_{ik} .
	\end{split}
\end{equation}
The number of Coop users can be computed accurately using (\ref{equ.bar_N_c_app}). However, the cardinality of $\Phi_\mathcal{N}$ is $K_0^{(K-1)B}$, and hence the computational complexity
exponentially increases with $K$. For example, when $K=K_0=15$ and $B=9$, we obtain $|\Phi_\mathcal{N}| = 1.35 \times10^{140}$.

In the sequel, we seek an alternative solution.

Noticing that the probability  that multiple \emph{hit file groups} exist simultaneously decreases with the growth of the number of {hit file groups}, we
only consider the case where only one or two hit file groups exist to approximate the average number of Coop users as follows. This approximation is accurate when the number of clusters is large.

The probability that a \emph{hit file group} only contains $\mathcal{G}_k$ can be obtained from (\ref{equ.P^h_k}) as $\prod_{j=1,j\neq k}^{K}(1-{(P^h_j)}^B){(P^h_k)}^{B}$. As a result, the probability of $n_{ik}=m$ ($1 \leq m \leq K$) when a
\emph{hit file group} only contains $\mathcal{G}_k$  is $C_{K}^{m}(P_{k})^m(1-P_{k})^{(K-m)}/P^h_k$. Then, the average number of Coop
users in the cases where only one hit file group exists can be derived as
\begin{equation}
	\label{N_c_appr_1}
	\begin{split}
		\bar{N^c_1} &= \sum_{k=1}^{K} \prod_{j=1,j\neq k}^{K} (1-{(P^h_j)}^B){(P^h_k)}^{B}\cdot\\ &B\sum_{m=1}^{K}\frac{C_{K}^{m}(P_{k})^m(1-P_{k})^{(K-m)}}{P^h_k}m\\
		&= \sum_{k=1}^{K} \prod_{j=1,j\neq k}^{K}(1-{(P^h_j)}^B){(P^h_k)}^{B-1}BP_kK.
	\end{split}
\end{equation}

The probability that there only exist two \emph{hit file groups}, $\mathcal{G}_{k_1}$ and $\mathcal{G}_{k_2}$, can be obtained from (\ref{equ.P^h_k}) as $\prod_{j=1,j\neq k_1,k_2}^{K}(1-{(P^h_j)}^B)({P^h_{k_1}})^{B}({P^h_{k_2}})^B$. Consequently, the
probability of $n_{ik_1}=m_1$ and $n_{ik_2}=m_2$ (where $1 \leq m_1 \leq K$, $1 \leq m_2 \leq K$ and $2 \leq m_1+m_2=m \leq K$) when the \emph{hit file groups} only contain $\mathcal{G}_{k_1}$ and $\mathcal{G}_{k_2}$  is $C_{m}^{m_1}C_{m-m_1}^{m_2}  \frac{{(p_{k_1})}^{m_1}{(p_{k_2})}^{m_2}}{P^h_{k_1}P^h_{k_2}}$. Then, the average number of Coop
users in the case where only two hit file groups exist can be derived as
\begin{equation}
	\label{N_c_appr_2}
	\begin{split}
		\bar{N^c_2} &=\sum_{\Phi_{k_1k_2}} \prod_{j=1,j\neq k_1,k_2}^{K}\left(1-{(P^h_j)}^B\right)({P^h_{k_1}})^{B}({P^h_{k_2}})^B \cdot\\
		&B  \sum_{m=2}^{K}\sum_{\Phi_{m_1m_2}} C_{m}^{m_1}C_{m-m_1}^{m_2}  \frac{{(p_{k_1})}^{m_1}{(p_{k_2})}^{m_2}}{P^h_{k_1}P^h_{k_2}} m\\
		&=\sum_{\Phi_{k_1k_2}}\prod_{k=1,k\neq i,j}^{K}\left(1-{(P^h_k)}^B\right)({P^h_i}{P^h_j})^{B-1} \cdot \\
		&B  \sum_{m=2}^{K}\sum_{\Phi_{m_1m_2}}\frac{mm!}{m_1!m_2!}{(p_{k_1})}^{m_1}{(p_{k_2})}^{m_2},
	\end{split}
\end{equation}
where $\Phi_{k_1k_2} = \{k_1,k_2|0\leq k_1,k_2\leq K,i\neq j\}$, $\Phi_{m_1m_2}=\{m_1,m_2|m_1+m_2=m\}$, whose  cardinality are  $K^2$ and $m^2$ ($m \leq K$), respectively.

By combining \eqref{N_c_appr_1} and \eqref{N_c_appr_2}, the average number of Coop users is approximated as
\begin{equation}
	\label{equ.bar_N_c2_app}
	\begin{split}
		&\bar{N^c} \approx \bar{N^c_1} +\bar{N^c_2},
	\end{split}
\end{equation}
which can be obtained much easier than \eqref{equ.bar_N_c_app}.

This proves Proposition \ref{p:3}.

\section{Proof of Proposition \ref{p:4}}
\label{a:4}

By simply subtracting $\bar{R^n_i}$ from $\bar{R^c_i}$ and considering $\bar{I}_i \gg \sigma^2$,
we can obtain that
\begin{equation}
	\label{equ.difference}
	\begin{split}
		\bar{R^c_i}-\bar{R^n_i} &= \mathbb{E}\{\log_2(1+\frac{P\sum_{j=1}^{B}r_{ij}^{-\alpha}}{B\sigma^2})\} \\ &-\mathbb{E}\{\log_2(1+\frac{Pr_{ii}^{-\alpha}}{\bar{I_i}})\} \nonumber \\
		& =\mathbb{E}\{\log_2(\frac{\bar{I_i}}{B\sigma^2})\}. \nonumber\\
	\end{split}
\end{equation}
When $\log_2(\frac{\bar{I_i}}{B\sigma^2})\geq 0$ (i.e., $\bar{I_i}\geq B\sigma^2$), $\bar{R^c_i}-\bar{R^n_i}\geq 0$.

This proves Proposition \ref{p:4}.

\bibliographystyle{IEEEtran}
\bibliography{CBQ_J}
\end{document}